\font\caps=cmcsc10 at 12pt
\font\eightrm=cmr8 at 8pt
\documentclass[12pt]{oxarticle}
\usepackage{amsmath, amssymb}

\newcommand{\LT}{\LaTeX}

\newcommand{\bt}{\begin{tabular}{c}}
\newcommand{\et}{\end{tabular}}

\newcommand{\eb}{\ee\be } 
\newcommand{\ebp}{\rt.\ee\be\lt.} 
\newcommand{\bmat}{\lt ( \begin{array} }
\newcommand{\emat}{  \end{array} \rt )}

\newcommand{\oM}{{\ov M}}

\newcommand{\ot}{{\ov t}}

\newcommand{\ED}{\end{document}}

\newcommand{\oy}{{\ov \y}}

\renewcommand{\a}{\alpha}	
\renewcommand{\b}{\beta}

\renewcommand{\d}{\delta}
\newcommand{\e}{\epsilon}
\newcommand{\ve}{\varepsilon}

\newcommand{\q}{\theta}

\newcommand{\y}{\psi}

\newcommand{\la}{\label}
\newcommand{\ci}{\cite}

\newcommand{\ds}{\documentstyle}	
\newcommand{\fr}{\frac}

\newcommand{\pa}{\partial}
\newcommand{\ov}{\overline}
\newcommand{\be}{\begin{equation}}
\newcommand{\ee}{\end{equation}}
\newcommand{\ba}{\begin{array}} 
\newcommand{\ea}{\end{array}}
\newcommand{\bea}{\begin{eqnarray}}
\newcommand{\eea}{\end{eqnarray}}
\newcommand{\ra}{\rightarrow}
\newcommand{\Ra}{\Rightarrow}

\newcommand{\lt}{\left}
\newcommand{\rt}{\right}

\newcommand{\ben}{\begin{enumerate}}
\newcommand{\een}{\end{enumerate}}
\newcommand{\bitem}{\begin{itemize}}
\newcommand{\eitem}{\end{itemize}}

\newcommand{\articlenumber}{\LT 3098}

\proofmodefalse
\begin{document}
\begin{center}

\vspace*{1in}
{\Huge A Note on\\
Symmetric Mass and Interaction Terms\\ for Weyl Spinors and SUSY\\[10pt] }
\renewcommand{\thefootnote}{\fnsymbol{footnote}}
\renewcommand{\thefootnote}{\arabic{footnote}}

{\caps John A. Dixon}\footnote{cybersusy@gmail.com}
\\[.3in] 
{\bf Abstract}
\end{center}
One can always write the mass matrix for Weyl spinors so that it is symmetric. However this is often not a good idea.  It is  usually incompatible with irreducibility of the fermion representations.  As a result, a symmetrized mass  term hides important symmetries and creates misleading difficulties that are not genuinely part of the  theory.  This is true for the Standard Model for example, and for its supersymmetric versions.  There is a related subtlety, involving symmetrization of the interaction terms, that is  central to the SUSY breaking mechanism of Cybersusy.
\\[0.1in]
\large

Many elementary particle physicists are now using Weyl spinors for practical calculations. This is a natural thing to do for supersymmetric theories, in particular.   A very useful, detailed  and comprehensive review of this notation, and the Feynman rules,  has appeared recently \ci{haberetal}.  There is also a related,  updated and clear introduction to SUSY, as currently understood, in \ci{martin}.   

\section{Bilinear Symmetric Mass Terms}

This Note takes exception to one aspect of  the assumptions used in \ci{haberetal}.    In Weyl notation one has a mass term that looks like
\be
m g_{ij}
\ve^{\a \b}
\y^i_{\a} 
\y^j_{\b} 
\la{reduciblemassterm}
\ee
and since the spinors anticommute, and the tensor $\ve^{\a \b}$ is antisymmetric,  this implies that the mass matrix is symmetric:
\be
g_{ij} = g_{ji}
\la{deceptivesymmetry}
\ee

The authors of  \ci{haberetal} base their analysis of mass terms on the assumption that the mass terms are symmetric as in (\ref{deceptivesymmetry}). In this Note we discuss how this complicates the analysis and usually leads to a lack of clarity. 

The point to be made arises from a simple and well known fact.  Interesting actions typically have more than one representation for the fermions.  For example the Standard Model (SM) and the Supersymmetric Standard Model (SSM) have many different irreducible representations tangled together in an intricate way.
When there is more than one irreducible representation the equation (\ref{deceptivesymmetry}) is not true unless one combines all the irreducible representations 
into one combined reducible representation. For a model like the SM  this would mean that we need to combine the left handed quarks, the right handed antiquarks, the left handed leptons and the right handed antileptons into one column spinor, which we can then call
 $\y^i_{\a}$, and then we can write the mass term as  (\ref{reduciblemassterm}).  But this spinor $\y^i_{\a}$ is not an eigenstate of Lepton or Baryon number, and it mixes the 
$SU(3) \times SU(2) \times U(1)$ representations in a silly way too.

To see what happens in more detail, let us consider the 
`up quark' mass term in the action:
\be
m t_{pq}
\ve^{\a \b}
\y^{cp}_{U \a} 
\y^{q}_{T c \b} 
\la{upquarkmassterm}
\ee
The complex conjugate is 
\be
m \ot^{pq}
\ve^{\dot \a \dot \b}
\oy_{U cp \dot  \a} 
\oy_{T q\dot  \b}^{c  } 
\la{upquarkmasstermcc}
\ee

Here $\y^{cp}_{U \a}$  represent the left handed up quarks.  They are in a $3$ of SU(3) colour, with  index $c=1,2,3$.  The spinor $\y^{q}_{T c \b} $ is the complex conjugate of the right handed up quarks, and they are in  a $\ov 3$ of SU(3) colour.  The indices $p,q=1,2,3$ are flavour indices and $\a,\b$ are Weyl spinor indices. The spinor $\y^{cp}_{U \a}$ is the top member of the left weak isospin doublet $\lt ( \begin{tabular}{c}
$\y^{cp}_{U \a} 
$\\
$\y^{cp}_{D \a} 
$\\
\end{tabular}
\rt )
$  and the spinor 
$\y^{p}_{T c\a} 
$ is a right isospin  singlet.   The equation 
\be
t_{pq}= t_{qp} {\rm \bf \; is\; not \; true\; in \; general,}
\ee
because there are two different spinors here, and so this is not of the form  (\ref{reduciblemassterm}). 
 So the  matrix $t_{pq}$ is generally a non-symmetric complex matrix in flavour space.  To get the mass matrix to be symmetric for the up quarks we would need to construct a reducible `row spinor' and a reducible `column spinor' like this
\be
\y^{p \a} \equiv 
\lt (
 \y^{cp \a}_{U }, 
 \;\; \y^{p\a}_{T d} 
\rt )
\ee
\be
\y^q_{\a} \equiv 
\lt (
\begin{tabular}{c}
$\y^{eq}_{U \a} 
$\\
$\y^{q}_{T f\a} 
$\\
\end{tabular}
\rt )
\ee
But this is not sensible, and it introduces confusion, because this `reducible spinor' combines spinors which have  different  quantum numbers  for  hypercharge U(1), for weak isospin SU(2), for colour SU(3), and for Baryon number.   The resulting `symmetric' matrix is of the form
\be
\y^{p \a}  g_{pq} \y^{q}_{ \a} \equiv 
\fr{1}{2} \y^{p \a} \lt (
\begin{tabular}{cc}
$0 $ & $t_{pq} \d^{f}_{c} $\\
$t_{qp} \d^{d}_{e}$ & $ 0 $\\
\end{tabular}
\rt )\y^{q}_{ \a} 
\la{artifice}
\ee
so that its `symmetry' is really only a redundant and artificial repetition.  This redundancy leads us to try to solve complications that are not really there.

As will appear below, one is better off to use the irreducible representations and to forget trying to force the mass matrix to be symmetric.   Writing the mass matrix for this case as a `symmetric, but redundant' matrix, as in  (\ref{artifice}), makes the situation seem much more complicated than it really is.
It is  easier if one  recognizes that the matrix $t_{pq}$ is generally a complex non-symmetric matrix.     

When one recognizes that the mass matrix is properly viewed as   complex and non-symmetric,  the diagonalization of the mass matrix,  and the analysis of the model,  are both simple.  They require no complicated matrix algebra such as 
is used in \ci{haberetal}. Here is how this works for the example of the up quarks above.  

The `left-handed' kinetic term has the form
\be
\y^{cs \a}_{U} \pa_{\a \dot \b} \oy_{U cs}^{ \dot \b}.
\la{lefthandedkin}
\ee
Consider the  transformation
\be
\y^{cs}_{U \a} \ra M^s_{Up}\y^{cp}_{U \a}.
\ee
Its complex conjugate is
\be
\oy_{U cs \dot \a} \ra \ov M^{\;\;\;\;\;q}_{Us}
\oy_{U cq \dot \a}.
\ee
Expression (\ref{lefthandedkin}) is invariant under this transformation 
provided that the transformation matrix is unitary:
\be
 M^s_{Up} \ov M^{\;\;\;\;\;q}_{Us}
= M^s_{Up} \lt ( M^{\dag} \rt )^{q}_{Us}
= \d^{q}_{p}
\ee
Similarly one has   the freedom to make the following transformation 
\be
\y^{s}_{Tc \a} \ra M^s_{Tp}\y^{p}_{T c\a},
\ee
with the  complex conjugate,
\be
\oy^c_{T s \dot \a} \ra \ov M^{\;\;\;\;\;q}_{Ts}
\; \oy^c_{T q \dot \a}
\ee
where the transformation matrix is unitary:
\be
 M^s_{Tp} \ov M^{\;\;\;\;\;q}_{Ts}
= M^s_{Tp} \lt ( M^{\dag} \rt )^{q}_{Ts}
= \d^{q}_{p}
\ee
This preserves the `right-handed' 
kinetic term of the form
\be
\y^{s \a}_{T c} \pa_{\a \dot \b} \oy_{T s}^{ c \dot \b}.
\la{righthandedkin}
\ee
The free quadratic Lagrangian for the up quarks consists of the mass term 
(\ref{upquarkmassterm}),  plus its complex conjugate 
(\ref{upquarkmasstermcc}),
plus the two kinetic terms  (\ref{lefthandedkin})  and  (\ref{righthandedkin}). The equations of motion immediately generate  two positive semi-definite hermitian mass matrices, namely
\be
H_{T q}^{p}=  \ov t^{sp}t_{sq} \equiv
  \lt ( t^{\dag}t \rt )_{\;\;q}^{p} 
\la{Therm}
\ee
and\footnote{The notation T in  $t^{T}$ denotes the 
transpose matrix,  and $\ov t^{ps}$ is the complex conjugate of $t_{ps}$.}
\be
H_{U q}^{p}=  \ov t^{ps}t_{qs}\equiv
  \lt (\ov t t^{T} \rt )_{\;\;q}^{p} 
\la{Uherm}
\ee
We can use the freedom of choosing the two unitary matrices $M_T$ and $M_U $ 
to diagonalize these two hermitian matrices $H_T$ and $H_U $ .   We  are guaranteed that these two matrices $H_T$ and $H_U $ have the same eigenvalues, because they are products of the same matrices $t$ and $\ov t$, although in a different order.

A related point is that the matrix $t_{qs}$ gets transformed by the two unitary matrix as follows:
\be
t_{st} \Ra t'_{pq} = M_{Up}^s t_{st}  M_{Tq}^t
\la{transformt}
\ee
and this new matrix satisfies
\be
t'_{sp} \ot^{'sq} = H_{T p}^{'q}
= M_{Ts}^q H_{T t}^{s}\oM_{Tp}^t \equiv \lt ( M_T H_T M_T^{\dag} \rt )^q_{\;p}
\ee
where $ H_{T p}^{'q}$ is a real diagonal positive semidefinite matrix whose diagonal terms are the mass-squared eigenvalues.  The equation for the $U$ matrices works the same way of course, and it has the same eigenvalues. The matrix 
$t'_{pq}$ is   complex and generally not symmetric.  This is of course  equivalent to the `Singular Value Decomposition' in section D.1 of \ci{haberetal}, where the original complex symmetric matrix needs to be transformed into a complex non-symmetric matrix en route to finding the mass eigenvalues and eigenvectors. But when we look at the irreducible factors, without enforcing the redundant symmetry, we get more insight into what is going on. 

Another advantage of using the irreducible representations  is that one can immediately  write down the Cabibbo-Kobayashi-Maskawa (CKM) matrix in a simple direct way.  It arises in the usual way from the left handed charged current that couples to the $W^+$ vector boson.  There is no reason for  the matrices $H_U$ and $H_{Dq}^p=
\ov b^{ps}b_{qs}$ to commute, and hence the  matrix $M_U M_D^{\dag}$ is generally not unity, because these two hermitian matrices $H_U$ and $H_{D}$ cannot be simultaneously diagonalized. It is not generally possible to reduce the unitary CKM matrix
\be
M^p_{Us} \lt ( M^{\dag} \rt )^s_{Dq}
\ee
to an orthogonal matrix, even using  the remaining freedom to choose three phases for each of the matrices $M_U$ and $M_{D}$.   As is well known, when there are three flavours, for the general case, there is one CP violating phase angle  remaining  even after all the freedom is used up.  

So we have demonstrated that it is better to use irreducible representations with non-symmetric complex mass terms, rather than to use reducible representations with symmetric complex mass terms.  This argument has been derived from an example with complex Dirac type mass terms.  These methods also apply to the following kinds of mass terms, which are also  discussed in
 \ci{haberetal}: 
\ben
\item
{\bf Pseudoreal representations with Dirac Masses:}  The existence of Dirac masses implies that there is a conserved quantum number like Baryon number, which entails separate left and right kinetic terms, as for the up quarks in (\ref{lefthandedkin}) and 
(\ref{righthandedkin}) above. For the pseudoreal  case  there is an additional antisymmetric matrix that occurs as a direct product in the mass matrix, and so one can deduce that the remaining part of the  mass matrix can be written as a complex antisymmetric matrix.  Again, as for the complex representations, the result is not irreducible.  The resulting  artificial and redundant antisymmetrization  introduces unnecessary complications, as did the symmetrization.
\item
{\bf Majorana masses:} 
 For Majorana mass terms there  is  typically  only one unitary matrix available, because there is only one kind of kinetic term, not two.  For Majorana masses there is  no conserved quantum number like Lepton or Baryon number.
So combining the relevant spinors into a column does not result in an artificial symmetry as in equation (\ref{artifice}) above. At first glance,   these mass terms  appear  to create a genuinely new problem, because a symmetric complex mass matrix is natural for Majorana mass terms.  

However, for the Majorana case, there is only one  hermitian matrix  to be diagonalized, not two.  Both of the relevant hermitian matrices 
(\ref{Therm}) and 
(\ref{Uherm}) are the same for a symmetric matrix 
$t_{(pq)}$.     We can use the one adjustable unitary matrix to diagonalize the one hermitian positive semi-definite mass matrix. The symmetric matrix $t_{pq}= t_{qp}$ for this case  remains symmetric under the transformation like (\ref{transformt}) because the two unitary matrices involved in the Majorana case have been reduced to one unitary matrix. 
\item
{\bf Mixed Majorana and Dirac Masses:} 
One can easily imagine  more  exotic situations with plenty of irreducible representations and mixed Majorana and Dirac masses.   For these it is still probably better to write out the action in detail and solve it explicitly using methods like those used above, rather than combine representations and write down an artificially symmetric mass term.   
\een
So in conclusion, for this section dealing with the mass terms, it seems fair to say that the discussion about  `Takagi diagonalization' in  \ci{haberetal} is unnecessarily complicated and not very transparent for the simpler cases with pure Dirac or Majorana masses. These more sophisticated methods,  like Takagi diagonalization, might be necessary and useful for exotic cases. At any rate, it is simplest to start trying to diagonalize the mass terms using the irreducible representations,  without combining the irreducible representations into reducible ones to achieve a redundantly symmetrized mass term.

\la{masssection}

\section{Trilinear Symmetric Interaction Terms}

Supersymmetric theories formulated using Weyl spinors encounter exactly the same issues as were reviewed in section \ref{masssection}.  Once again it  appears that the superpotential $P$ has a symmetric mass term $g_{ij}=g_{ji}  $ and also a symmetric interaction term $g_{ijk}=g_{(ijk)}$:
\be
P = \int d^4 x d^2 \q 
\lt \{
m^2 g_{i} {\hat A}^i + m g_{ij} {\hat A}^i {\hat A}^j 
+  g_{ijk} {\hat A}^i {\hat A}^j {\hat A}^k  \rt \}
\la{superstuff}
\ee
However this is generally not   a good way\footnote{Both the symmetric notation and the non-symmetric are useful in different contexts.   The symmetric notation is useful for general considerations, as illustrated in equations (\ref{constraint}), 
(\ref{eqpotent3}) and (\ref{potent3}). However, when one looks at models like the Supersymmetric Standard Model (SSM), this should often be done with the non-symmetric notation. For example, both kinds of notation are used to simplify  the discussion in \ci{susy09}.} to look at models like the  SSM.  It is   better  to use non-symmetric matrices $g_{ij}$  and also non-symmetric interaction terms $g_{ijk}$.  The natural notation uses irreducible representations of the symmetries, which depends on the quantum numbers, and it contains no artificial symmetrization like in equation (\ref{artifice}). 

  Of course, if one breaks supersymmetry explicitly, this introduces complications, but generally it is still useful to regard the mass matrices as complex and non-symmetric.

These symmetrization issues are more subtle than they appear to be. There is a folkloric tendency to believe that the unified theory of everything must be based on an irreducible model with one huge representation of some huge group.  This results from an aesthetic notion that irreducibility is equivalent to simplicity.  However, it turns out  that   certain combinations of irreducible representations can mingle with each other through the constraint equations for SUSY.  This issue goes right to the heart of the supersymmetry breaking mechanism of Cybersusy, and we will see that the SSM is quite special.  

Cybersusy \ci{susy09} has a constraint equation which looks like this
\be
{\cal L}_{f}  P_3 
=0
\la{constraint}
\ee
where ${\cal L}_f$ is a Lie algebra generator, made of scalars, of the form
\be
{\cal L}_{f}  = f_{i}^{j} A^i \fr{\pa}{\pa A^j}  
\la{eqpotent3}
\ee
and $P_3$ is the unintegrated trilinear scalar part of the superpotential, extracted from (\ref{superstuff}):
\be
P_3 = g_{pqr} A^p A^q A^r
\la{potent3}
\ee
These equations are written in the symmetric form, and they typically involve redundant and artificial symmetrization, as discussed above.  The difference is that the redundant and artificial symmetrization for this aspect of Cybersusy relates to the trilinear interaction term, not the bilinear mass term.  I was prevented for years from finding physically interesting solutions for (\ref{constraint}) 
  by this very issue of artificial and redundant symmetrization.  I did not realize that if one tries to solve the relevant equations in the symmetric form above, one never finds the following solutions, because they are intrinsically non-symmetric, just as the mass matrices for the SM and SSM are.

Physically interesting solutions of this equation arise when one writes these equations in the non-symmetric form, using the non-symmetric form of the superpotential for the SSM, as follows.  The scalar part of the trilinear term of the superpotential for the SSM\footnote{This is a non-minimal version of the SSM. It has   right handed neutrinos $R^p$ with Dirac type masses for the neutrinos, and it also has a singlet Higgs field J designed to spontaneously break $SU(2) \times U(1) $ down to $U(1)$, when one includes a term $-m^2 J$ in the potential.  Note that this $J$ singlet also plays an important role in the Lie algebra generators.}   has the following non-symmetric form:

\[
P_{{\rm SSM} }
=
g \ve_{ij} H^i K^j J
+
p_{p q} \ve_{ij} L^{p i} H^j P^{ q} 
+
r_{p q} \ve_{ij} L^{p i} K^j R^{ q}
\]
\be
+
t_{p  q} \ve_{ij} Q^{c p i} K^j T_c^{ q}
+
b_{p  q} \ve_{ij} Q^{c p i} H^j B_c^{ q}
\la{PSSM}
 \ee
This notation is closely related to the notation used in equation (\ref{upquarkmassterm}) above. 
Here the fields $J,H^i,K^i$ are Higgs/Goldstone  scalar fields from the respective supermultiplets, with hypercharge $Y=0,-1,+1$ respectively.  $ Q^{c p i}$ is the scalar from the Left Quark supermultiplet with hypercharge $Y=\fr{1}{3}$. $T_c^{ q}$ and $B_c^{ q}$
are the scalars from the right handed up and down antiquark supermultiplets with hypercharge 
$Y=-\fr{4}{3}, \fr{2}{3}$ respectively.  
$ L^{ p i}$ is the scalar from the Left Lepton supermultiplet with hypercharge $Y=-1$.  $P^{ q} $ and   $R^{ q} $ are the scalars from the right handed antipositron and antineutrino  supermultiplets, with hypercharge $Y=2,0$ respectively. The indices $i,j=1,2$ are weak SU(2) indices. 

Then the following physically interesting Lie algebra operators exist
for the Leptons:
\be
{\cal L}^{pi}_{L}
=
g^{-1} L^{p i} \fr{\pa}{\pa J}
+
(p^{-1})^{qp} K^{ i} \fr{\pa}{\pa P^q}
-
(r^{-1})^{qp} H^{ i} \fr{\pa}{\pa R^q}
\la{thetrickyoneforleptons}
\ee
\be
{\cal L}^p_{P}
=
g^{-1} P^{p } \fr{\pa}{\pa J}
+
(p^{-1})^{pq} K^{ i} \fr{\pa}{\pa L^{iq}}
\la{rightelect}
\ee
\be
{\cal L}^p_{R}
=
g^{-1} R^{p } \fr{\pa}{\pa J}
-
(r^{-1})^{pq} H^{ i} \fr{\pa}{\pa L^{iq}}
\ee
where  the inverse matrices are defined in the following way:
\be
p_{s q}   (p^{-1})^{qp} 
=(p^{-1})^{pq}  
p_{q s}=
\d^{p}_{s};
\ee
\be
r_{s q}   (r^{-1})^{qp} 
=
   (r^{-1})^{pq} r_{q s}
=
\d^{p}_{s}.
\ee
Similarly,  the following  physically interesting Lie algebra operators exist
for the Quarks:
\be
{\cal L}^{cpi}_{Q}
=
g^{-1} Q^{c p i} \fr{\pa}{\pa J}
-
(t^{-1})^{qp} H^{ i} \fr{\pa}{\pa T^q_c}
+
(b^{-1})^{qp} K^{ i} \fr{\pa}{\pa B^q_c}
\la{thetrickyoneforquarks}
\ee
\be
{\cal L}^p_{T c}
=
g^{-1} T^p_c \fr{\pa}{\pa J}
-
(t^{-1})^{pq} H^{ i} \fr{\pa}{\pa Q^{icq}}
\ee
\be
{\cal L}^p_{Bc}
=
g^{-1} B^p_c \fr{\pa}{\pa J}
+
(b^{-1})^{pq} K^{ i} \fr{\pa}{\pa Q^{icq}}
\la{rdquark}
\ee
where  the inverse matrices are defined in the following way:
\be
 (t^{-1})^{pq} t_{qs}
=
 t_{sq}(t^{-1})^{qp} 
=
\d^{p}_{s}
\ee
\be
(b^{-1})^{pq} b_{qs}
=
 b_{sq}(b^{-1})^{qp}
=
\d^{p}_{s}
\ee
Using the above forms  
(\ref{PSSM}) and 
(\ref{thetrickyoneforleptons}), for example,  
it is easy to verify that:
\be
{\cal L}^{pi}_{L}
P_{{\rm SSM} }
=0
\ee
as follows:
\be
{\cal L}^{pi}_{L}
P_{{\rm SSM} }
\eb
=
\lt \{
g^{-1} L^{p i}g \e_{jk} H^j K^k  
+
(p^{-1})^{qp} K^{ i} p_{s q} \e_{jk} L^{s j} H^k
-
(r^{-1})^{qp} H^{ i} r_{s q} \e_{jk} L^{s j} K^k
\rt \}
\eb
=
\lt \{
  L^{p i}  \e_{jk} H^j K^k  
+
     \e_{jk} L^{p j}
\lt ( K^{ i} H^k
-
 H^{ i}   K^k
\rt )
\rt \}
\ee

Now use
\be
 K^{ i}     H^k
- H^{ i}     K^k
= \ve^{ik} \lt ( \ve_{lm}K^{l} H^m
\rt )
\ee
and we get 

\be
{\cal L}^{pi}_{L}
P_{{\rm SSM} }
=
\lt \{
  L^{p i}  \e_{jk} H^j K^k  
+
     \e_{jk} L^{p j}
\ve^{ik} \lt ( \ve_{lm}K^{l} H^m
\rt )
\rt \}
=0
\ee
We also have, using (\ref{PSSM}) and (\ref{rightelect}), 

\be
{\cal L}^p_{P}P_{{\rm SSM} }
=
\eb
g^{-1} P^{p } g \e_{ij} H^i K^j
+
(p^{-1})^{pq} K^{ i} \lt ( 
p_{q s} \e_{ij}   H^j P^{ s} 
+
r_{q s} \e_{ij}   K^j R^{ s}
\rt )
\ee
Now observe that
\be
K^{ i} 
 \e_{ij}   K^j 
=0
\ee
So we get
\be
{\cal L}^p_{P}P_{{\rm SSM} }
=
\eb
 P^{p }  \e_{ij} H^i K^j
+
 K^{ i}  
  \e_{ij}   H^j P^{ p} 
=0
\ee

The other four Lie algebra operators work in a similar way. 
Observe the intertwining of the left doublets and right singlets here, and the crucial role of the singlet and the two SU(2) Higgs doublets. This all seems quite specific to the SSM.  It is not obvious whether other models have physically interesting solutions too.  In the SSM, 
each of these six Lie algebra invariance generators\footnote{When the gauge symmetry breaks down $SU(2)\times U(1)\ra U(1)$, these develop the Cybersusy algebra in a way that looks very promising for the elimination of unnaturally large flavour changing neutral currents, because the SUSY breaking is naturally flavour diagonal as between scalars and spinors. I had not realized this when these generators were first written down in a less transparent and partially incorrect  form in \ci{susy09}. }  has a Lepton (Quark) scalar multiplied by $\fr{\pa}{\pa J}$, added to terms made from the Higgs $H^i,K^i$ multiplied by the derivative of an  Antilepton (Antiquark) scalar  (or vice versa).  Each of them is in a representation  of the gauge groups $U(1)$, $SU(2)$ and $SU(3)$.  Each of them is in an eigenstate of Quark and Lepton number.  These six operators form an invariant Abelian subalgebra of the invariances of the term (\ref{PSSM}). This invariance algebra includes the generators of $SU(3) \times SU(2) \times U(1)$ as well as Baryon and Lepton number.

These invariances would not exist if the standard model did not have its peculiar left-right asymmetry, which is also carried through to the Higgs sector in this supersymmetric version of the SM.  Also note that this invariance is far from obvious if one writes the superpotential in an artificially symmetrized and reducible way.

These Lie algebra operators are central to Cybersusy\footnote{The five long papers on Cybersusy in arXiv are somewhat out of date now, because of progress like that introduced in this Note.} and its mechanism\footnote{SUSY breaking for other sectors, such as the Higgs/Gauge sector, is still in the exploratory stage but it looks promising.} for supersymmetry breaking.  In \ci{susy09}, I had not yet realized that the operators 
(\ref{thetrickyoneforleptons}) and 
(\ref{thetrickyoneforquarks}) existed, and the paper \ci{susy09} asserts that cybersusy only works for the right handed quarks and leptons in the SSM.  In fact it works for both left and right quarks and leptons, because (\ref{thetrickyoneforleptons}) and 
(\ref{thetrickyoneforquarks}) do exist. 

In conclusion, it is necessary to be very careful when using  symmetrized matrices $g_{(ij)}$ and tensors  $g_{(ijk)}$ for Weyl spinors and SUSY, because the symmetrization forces one to use reducible and redundant representations.   It is often better to write the action in terms of irreducible components,  even though the matrices  $g_{ij}$ and  tensors $g_{ijk}$  are  usually  not symmetric when written in that way. Use of the matrices $g_{(ij)}$ and tensors  $g_{(ijk)}$  with artificial or redundant symmetry 
 makes  the diagonalization of mass matrices, and  the solution of the Cybersusy constraint equations, appear to be more difficult than they are.

\end{document}